\documentclass[12pt]{article}
\usepackage{graphicx}
\usepackage{color}
\include{feynman}

\def\beq{\begin{equation}}
\def\eeq#1{\label{#1}\end{equation}}
\def\eeqn{\end{equation}}

\def\beqa{\begin{eqnarray}}
\def\eeqa#1{\label{#1}\end{eqnarray}}
\def\eeqan{\end{eqnarray}}

\let\bar=\overbar

\def\Dslash{\not{\hbox{\kern-4pt $D$}}}
\def\dslash{\not{\hbox{\kern-2pt $\del$}}}

\def\msb{{\bar{\ssstyle M \kern -1pt S}}}

\def\Title#1{\begin{center} {\Large {\bf #1} } \end{center}}

\begin{document}

\rightline{\normalsize UASLP--IF--07--001}
\rightline{\normalsize FERMILAB--Conf--07/029--E}
\Title{SELEX: Recent Progress in the Analysis
of Charm-Strange and Double-Charm Baryons}

\begin{center}{\large \bf Contribution to the proceedings of HQL06,\\
Munich, October 16th-20th 2006}\end{center}

\bigskip\bigskip


\begin{raggedright}  

{\it J\"urgen Engelfried\footnote{email: {\tt jurgen@ifisica.uaslp.mx}}
\index{Engelfried, J.}\\
Instituto de F\'{\i}sica,\\
Universidad Aut\'onoma de San Luis Potos\'{\i}, Mexico\\
for the SELEX Collaboration}
\bigskip\bigskip
\end{raggedright}

\section{Introduction}
SELEX (Fermilab Experiment 781)~\cite{inst}
employs beams of $\Sigma^-$, $\pi^-$, and
protons at around $600\,\mbox{GeV}/c$ to study production and decay properties
of charmed baryons.  It took data in the 1996/7 fixed target run and is
currently analyzing those data.

Here we will focus on recently obtained results concerning
the $\Omega_c^0$ lifetime and the doubly-charmed baryons $\Xi_{cc}^+$ 
and $\Xi_{cc}^{++}$.

\section{New Results on the {\mbox{\boldmath$\Omega_c^0$}}}

\begin{figure}[ht]
\begin{center}
\includegraphics[width=0.32\textwidth]{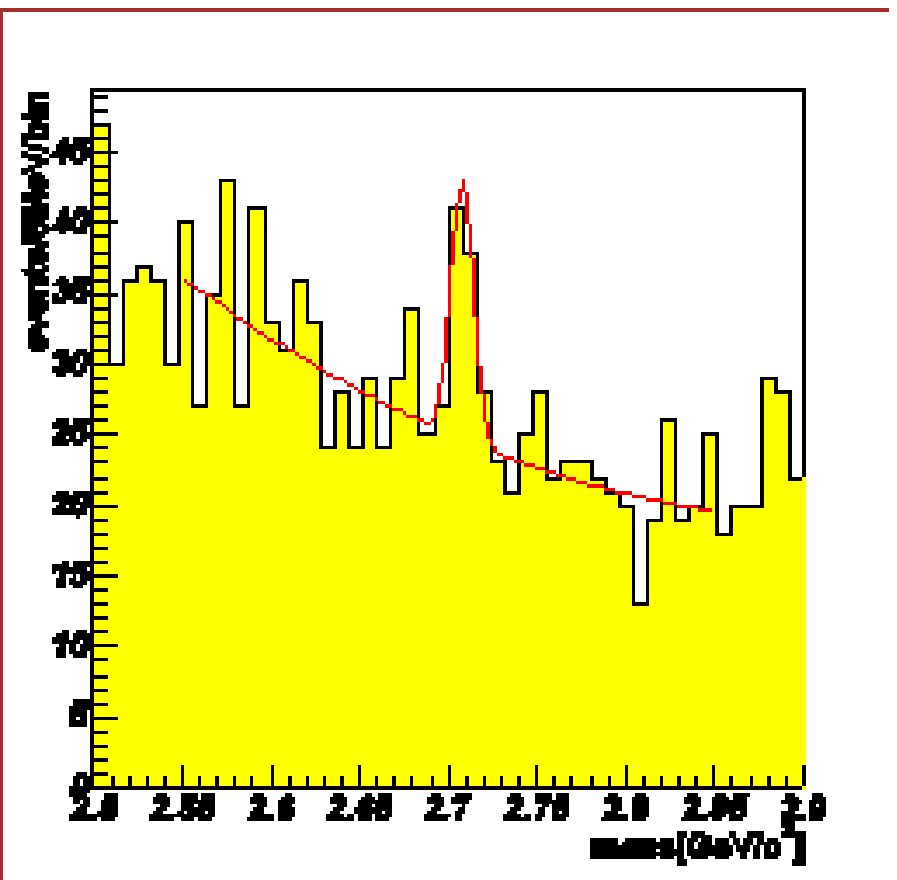}
\includegraphics[width=0.32\textwidth]{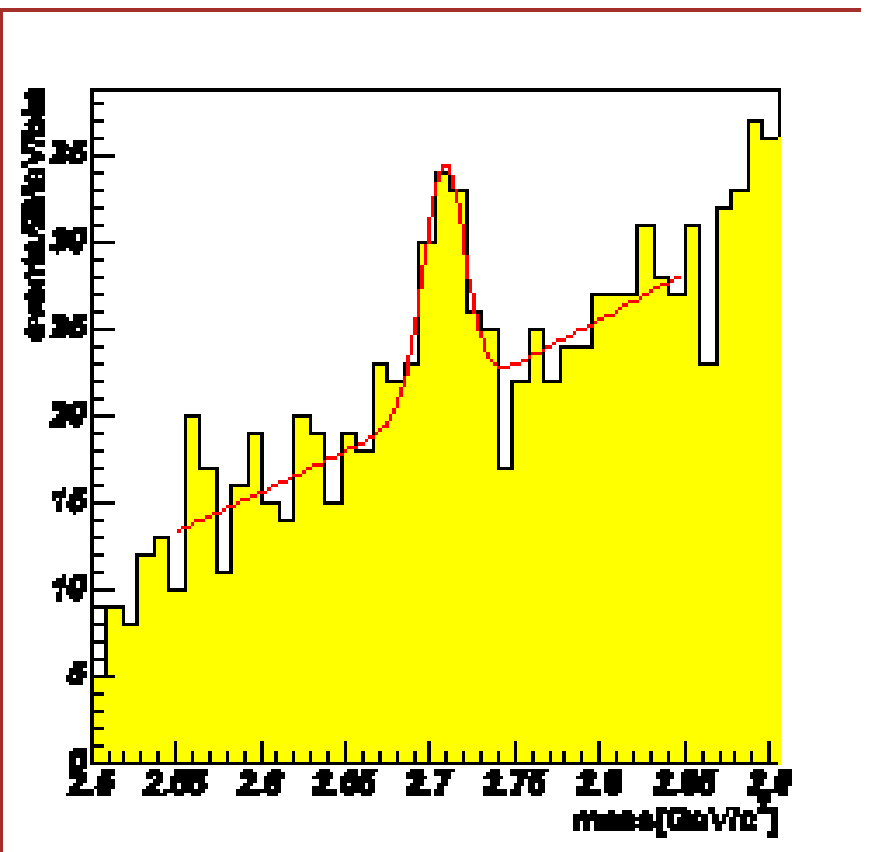}
\includegraphics[width=0.32\textwidth]{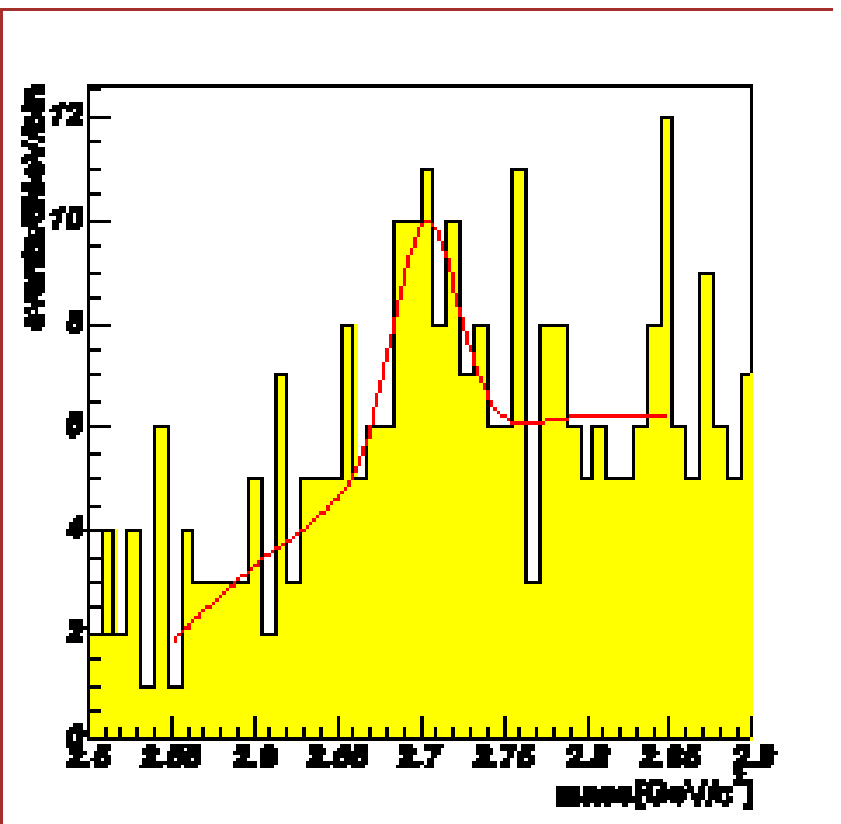}
\end{center}
\caption{
Invariant mass distributions for different decay modes of
the $\Omega_c^0$. 
Left: $\Omega^-\pi^+$, Signal: $35\pm12$ events;
center: $\Omega^-\pi^+\pi^+\pi^-$, $44\pm14$ events;
right: $\Xi^-K^-\pi^+\pi^+$, $28\pm12$ events
}
\label{omcmass}
\end{figure}

SELEX observes the $\Omega_c^0$ in three decay modes, namely
$\Omega_c^0\to\Omega^-\pi^+$, 
$\Omega_c^0\to\Omega^-\pi^+\pi^+\pi^-$, and
$\Omega_c^0\to\Xi^-K^-\pi^+\pi^+$. The invariant mass distributions
for these modes are shown in fig.~\ref{omcmass}.
The total sample contains $107\pm22$ events, 
nearly half of them in $\Omega3\pi$.  At this moment we are working on the
systematics of the mass and branching ratio measurements of these
modes~\cite{sedat}.

We use the $\Omega\pi$ and $\Omega3\pi$ channels to determine the
lifetime of the $\Omega_c^0$.
We calculate the reduced proper time $ct$, given by 
$ct=(L-N\sigma)/\gamma$, requiring $L/\sigma>N$ with $N=6$, for each
event within the mass region of the $\Omega_c^0$. The proper lifetime
resolution is $\sim 20\,\mbox{fs}$. 
We make a maximum likelihood fit to
a probability distribution having an exponential decay
for the signal and two exponentials for the fast and slow
components of the background:
\begin{displaymath}
N_s(1-\alpha)f(t)\tau^{-1} e^{-t/\tau} +
\alpha N_B(\beta\tau_1^{-1} e^{-t/\tau_1} +
(1-\beta)\tau_2^{-1} e^{-t/\tau_2})
\end{displaymath}
where $\tau$, $\alpha$, $\beta$, $\tau_1$, $\tau_2$
are the fit parameters describing the lifetime and the relative 
contributions of the background to the $ct$ distribution, and
$f(t)$ is the acceptance function.
We do this for each mode separately, and obtain
for the $\Omega\pi$ mode $\tau=62.6 \pm 22.0\,\mbox{fs}$ and for the
$\Omega3\pi$ mode $\tau=65.8 \pm 16.0\,\mbox{fs}$. Combining the two
results yields
$\tau_{\Omega_c}=65 \pm 13\,(stat) \pm 9\,(sys)\,\mbox{fs}$.
More details can be found in~\cite{omclife}.
This result should be compared to the current 
PDG average~\cite{pdg}
of $69\pm12\,\mbox{fs}$, using a total of 175~events
from three different experiments.

\section{Doubly Charmed Baryons}
\subsection{The Discovery of Double Charm Baryons}

In 2002 the SELEX collaboration reported the first observation of a
candidate for a double charm baryon, 
decaying as $\Lambda_{c}^+ K^-\pi^+$~\cite{prl,Thesis}.  The
state had a mass of $3519\pm2\,\mbox{MeV}/c^2$,
and its observed width was consistent
with experimental resolution, less than $5\,\mbox{MeV}/c^2$.  The final state
contained a charmed baryon and negative strangeness ($\Lambda_c^+$ and $K^-$),
consistent with the Cabibbo-allowed decay of a $\Xi_{cc}^+$ configuration. 
In order to confirm the interpretation of this state as a double charm
baryon, it is essential to observe the same state in some other way.
Other experiments with large charm baryon samples, e.g., the FOCUS~\cite{FOCUS}
and E791 fixed target charm experiments at Fermilab or the 
B-factories,
have not confirmed the double charm signal.  This is not inconsistent
with the SELEX results. The report in Ref.~\cite{prl} emphasized that 
this new state was produced by the baryon beams ($\Sigma^-$, proton) in 
SELEX, but not by the $\pi^-$ beam.  It also noted that the apparent lifetime
of the state was significantly shorter than that of the $\Lambda_c^+$, which
was not expected in model calculations~\cite{guberina}.
A more detailed discussion can be found in~\cite{goteborg}.

\subsection{Features and Problems in the Original Analysis,
and Possible Solutions}

All the signals observed so far are statistically significant, 
but have only a few signal events.
The signals are clean, e.g.\ there is very little
background, but the background itself is also difficult to estimate.
SELEX only observes events from the baryon ($\Sigma^-$, proton) beams,
and the number of observed events is larger than some
production models (see for
example~\cite{murray,kiselev}) predict. As mentioned before, the lifetime seems
to be very short, and no other experiment has confirmed our observations.

Another way to confirm the $\Xi_{cc}^+$ is to observe
it in a different decay mode that also involves a final state with baryon
number and charm (not anti-charm).  One such mode involving only stable charged
particles is the channel $pD^+K^-$,
another one $\Xi_c^+\pi^+\pi^-$. SELEX developed a new method for
a more reliable background determination.  We also improved the resolution
on the secondary vertex position by including the single-charm track into
the vertex fit, and we redid our full analysis chain to increase our
statistics.  In the following we will describe these step in details.

\subsection{New Analysis Features within SELEX}
The Cabibbo-allowed decay of the $\Xi_{cc}^+$ is shown in the following figure.
\begin{center}
\begin{picture}(20000,15000)
\THICKLINES
\drawline\fermion[\E\REG](1000,3000)[15000]
\drawline\photon[\NE\REG](\particlemidx,\particlemidy)[6]
\drawline\fermion[\N\REG](\photonbackx,\photonbacky)[3000]
\put(\particlebackx,\particlebacky){$u$}
\global\Xthree=\particlebackx
\global\advance\Xthree by 500
\global\Ythree=\particlebacky
\global\advance\Ythree by 400
\drawline\fermion[\E\REG](\photonbackx,\photonbacky)[3000]
\put(\particlebackx,\particlebacky){$\overline{d}$}
\put(0,2600){$c$}
\put(16500,2600){$s$}
\put(8200,5000){$W^+$}
\drawline\fermion[\E\REG](1000,0)[15000]
\put(0,-400){$d$}
\put(16500,-400){$d$}
\drawline\fermion[\E\REG](1000,1500)[15000]
\put(0,1100){$c$}
\put(16500,1100){$c$}
\end{picture}
\end{center}
In the final state we expect a baryon, 
and the quarks $csdu\bar{d}$ plus eventually some pairs from the sea.
We also expect a cascaded decay chain, with the first, and later the
second charm quark undergoing a weak decay.

For SELEX, the easily accessible decay modes for the different 
doubly charmed baryons are:
$\Xi_{cc}^+\to \Lambda_c^+ K^- \pi^+$,
$\Xi_{cc}^+\to p D^+ K^-$,
$\Xi_{cc}^+\to \Xi_c^+ \pi^- \pi^+$,
$\Xi_{cc}^{++}\to \Lambda_c^+ K^- \pi^+\pi^+$,
$\Xi_{cc}^{++}\to p D^+ K^-\pi^+$
(depending on the mass of the $\Xi_{cc}^{++}$),
$\Xi_{cc}^{++}\to \Xi_c^+ \pi^+$,
$\Xi_{cc}^{++}\to \Xi_c^+ \pi^+ \pi^+\pi^-$,
$\Omega_{cc}^+\to \Xi_c^+ K^- \pi^+$, and
$\Omega_{cc}^+\to \Xi_c^+ K^- \pi^+\pi^+\pi^-$.
The first two modes are already published~\cite{prl,SELEX2} by SELEX, and
work on the other modes is in progress; here we will report on a first
observation of the third decay mode listed.

For the background determination, we developed an event mixing method.
The first decay vertex is close to the primary vertex, and we assume 
that all the background is purely combinatoric.  We make combinatoric
backgrounds by taking the first decay vertex from one event, and the second
vertex from another event; to increase statistics, we use the single-charm
vertex 25~times.  The resulting combinatoric background is absolutely
normalized. We employed this method already in~\cite{SELEX2}.

\subsection{{\mbox{\boldmath$\Xi_{cc}^+\to \Lambda_c^+ K^- \pi^+$}} -- 
New Analysis}
\begin{figure}[ht]
\hfill
\includegraphics[clip,width=6cm]{lc-old.eps.fixed}
\hfill
\includegraphics[width=8.0cm]{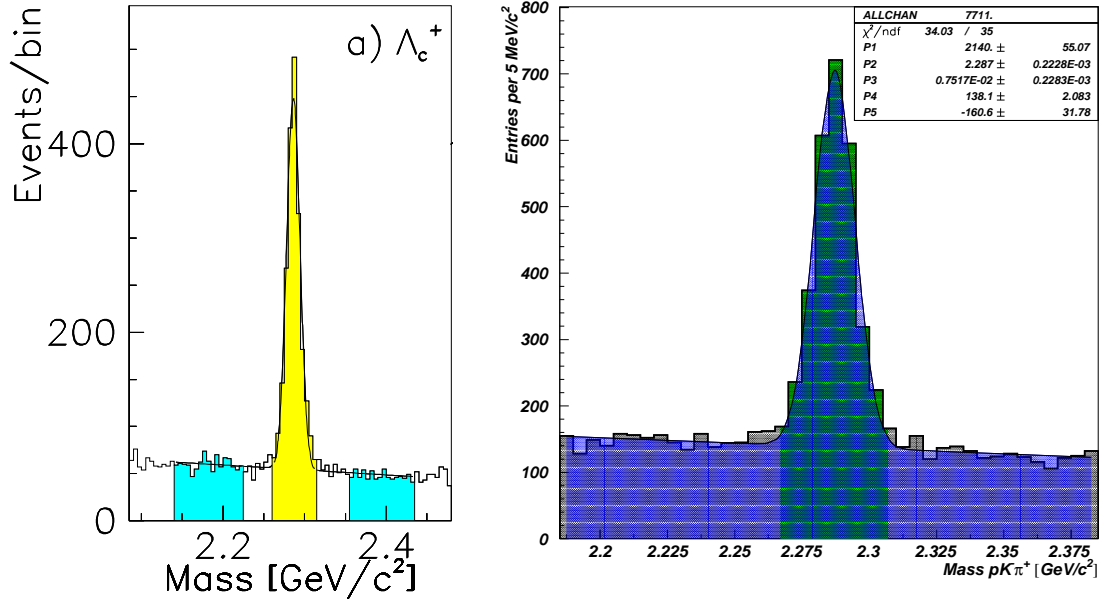}
\hfill
\caption{$\Lambda_c^+\to pK^-\pi^+$ data sets of original (left) and new
(right) analysis.}
\label{lamc}
\end{figure}
To increase our statistics, we re-analyzed our full data set with some
softer cuts and with improved tracking software. In fig.~\ref{lamc}
we show a comparison of the $\Lambda_c^+$ data set used for the
analysis.
The number of $\Lambda_c\to pK^-\pi^+$ candidates increased from 1630 to 2140.

We also improved the resolution of the decay vertex position of 
the $\Xi_{cc}^+$ candidate
by including the vector of the $\Lambda_c^+$ into the
vertex fit. This improved resolution reduces the background when applying 
a cut in $L/\sigma$, while keeping more signal events. It also
increases the possibility of measuring the lifetime of double charm baryons.

\begin{figure}[ht]
 \centerline{\includegraphics[width=0.8\textwidth]{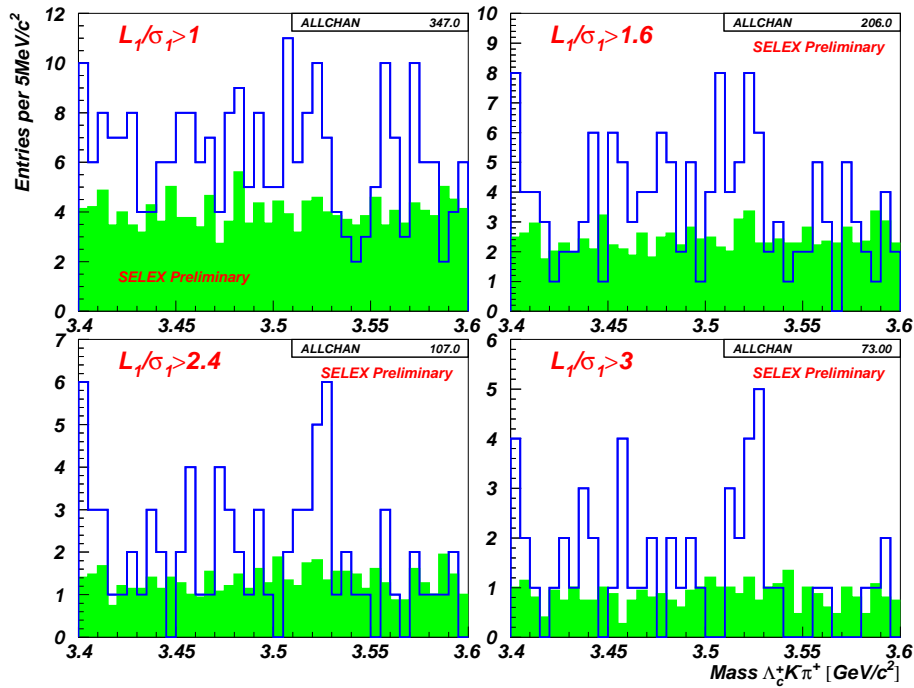}}
\caption{$\Lambda_c^+K^-\pi^+$ ($\Lambda_c^+\to pK^-\pi^+$)
 invariant mass distributions (blue) for
various cuts in $L/\sigma$ on the first decay vertex. In green we show
the estimated combinatoric background from the event mixing procedure
described in the text.}
\label{lcls}
\end{figure}
Figure~\ref{lcls} shows the results of our new analysis, for various
cuts in $L/\sigma$ of the first decay vertex.
Re-analyzing and relaxing some cuts in the single charm sample
increased the number of signal events, but also
resulted in a somewhat higher background level;
but the background is nicely reproduced and well understood from the
combinatoric analysis. The improved secondary vertex resolution yields in
cleaner signals and allows access to other decay modes, which we will 
pursue in the future.  Measuring the lifetime now seems possible, but is
still challenging. As seen from the yields for different cuts in $L/\sigma$,
the lifetime seems to be around $1\,\sigma$.

\subsection{{\mbox{\boldmath$\Xi_{cc}(3780)^{++}\to\Lambda_c^+K^-\pi^+\pi^+$}}}
\begin{figure}[ht]
\centerline{\includegraphics[width=0.6\textwidth]{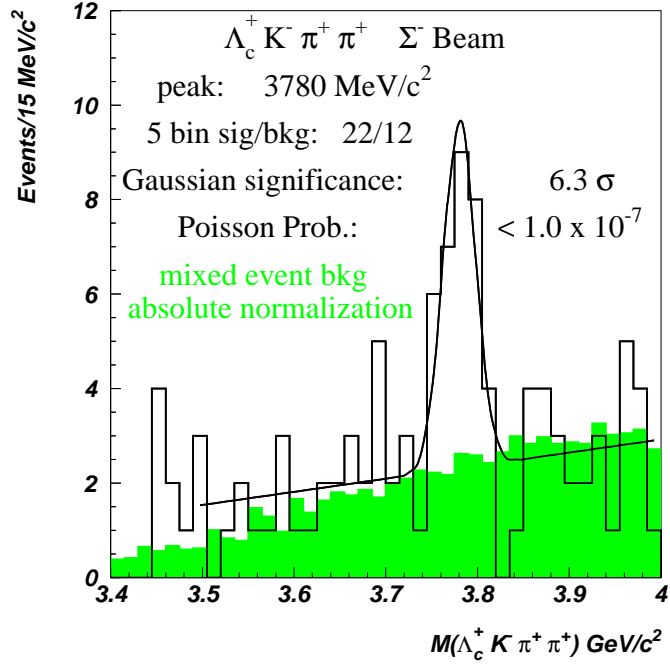}}
\caption{The $\Lambda_c^+K^-\pi^+\pi^+$ invariant mass distribution,
for $\Sigma^-$ beam only.}
\label{xicc3780}
\end{figure}
We also revisited with our re-analyzed data set the first double-charm
baryon state we found in SELEX~\cite{Thesis}, the $\Xi_{cc}(3780)^{++}$.
In fig.~\ref{xicc3780} is shown the $\Lambda_c^+K^-\pi^+\pi^+$
invariant mass distribution, restricting ourselves to $\Sigma^-$ induced
events.  The peak at $3780\,\mbox{MeV}/c^2$ is statistically significant,
and is wider than our experimental resolution, as shown by Monte Carlo.
The background is well described by our mixed event procedure. By removing
the slower of the $\pi^+$'s, we observe that about half of the 
$\Xi_{cc}(3780)^{++}$ decay to $\Xi_{cc}^+(3520)$.  At this moment we
are finishing up the analysis for this state.

\clearpage
\section{First Observation of
{\mbox{\boldmath$\Xi_{cc}^+\to \Xi_c^+ \pi^+ \pi^-$}}}
\begin{figure}[ht]
\centerline{\includegraphics[width=0.8\textwidth]{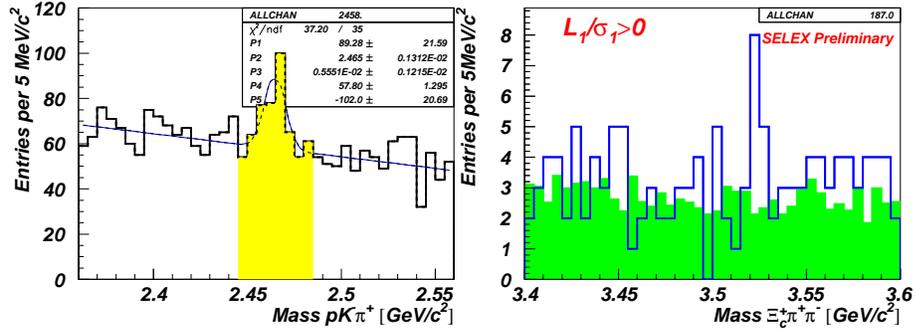}}
\caption{Left: $pK^-\pi^+$ invariant distribution and $\Xi_c^+$ sample
(yellow) used. Right: $\Xi_c^+ \pi^+ \pi^-$ invariant mass distribution.
The green histogram is our estimate of the combinatoric background.}
\label{xiccc}
\end{figure}

SELEX published~\cite{sun} the first observation of the
Cabibbo-suppressed decay
of $\Xi_c^+\to pK^-\pi^+$; this is the same final state as we used
before for the reconstruction of the $\Lambda_c^+$. Our sample of
$\Xi_c^+$ in the mode is  much smaller than our $\Lambda_c^+$ sample,
but the branching fraction of $\Xi_{cc}^+\to\Xi_c^+\pi^+\pi^-$ should
be larger than to $\Lambda_c^+K^-\pi^+$.  We applied the same cuts
and procedure as to the previously described analyzes, and
obtained~\cite{ibrahim}
the $\Xi_c^+\pi^+\pi^-$ invariant mass distribution shown in
fig.~\ref{xiccc}.
A clear peak at about $3520\,\mbox{MeV}/c^2$ is seen in the figure. This
constitutes the first observation of this decay mode of the $\Xi_{cc}^+(3520)$.

\section{Summary}
SELEX is still the only experiment observing double charm baryons. We
published observations on two different decays modes,
$\Xi_{cc}^+\to \Lambda_c^+ K^- \pi^+$~\cite{prl} and
$\Xi_{cc}^+\to pD^+K^-$~\cite{SELEX2}.
After a re-analysis of our full data set, with improved efficiency
and resolution, we presented here a higher-statistics observation of
$\Xi_{cc}^+\to \Lambda_c^+ K^- \pi^+$, and a re-analysis of the
$\Xi_{cc}(3780)^{++}$. The new analysis also allows access to additional
decay modes, and we presented here the
first observation of $\Xi_{cc}^+\to \Xi_c^+ \pi^- \pi^+$.

SELEX will continue the line of analysis, by first publishing these
preliminary results.  We will try to measure the lifetime of the $\Xi_{cc}^+$.
We will also seek the isospin-partner of the $\Xi_{cc}^+$, the $\Xi_{cc}^{++}$
in all corresponding decay modes around $3500\,\mbox{MeV}/c$.

\section{Acknowledgment}
The author thanks the organizers for the invitation to present these
results at the conference.  This work was supported in part by
the Consejo Nacional de Ciencia y Tecnolog\'{\i}a (CONACyT), Mexico, 
and by a special research grant of UASLP.

\end{document}